\documentclass[aps,prc,preprint,groupedaddress]{revtex4-1}

\usepackage{graphicx}

\usepackage{amsmath}
\usepackage{amssymb}
\def\bm{\boldsymbol}
\newcommand{\bea}{\begin{eqnarray}}
\newcommand{\eea}{\end{eqnarray}}
\newcommand{\be}{\begin{eqnarray}}
\newcommand{\ee}{\end{eqnarray}}
\newcommand{\no}{\nonumber \\}

\def\vp{{\bm p}}

\def\vs{{\bm\sigma}}

\begin{document}




\title{Symmetry violations in few-body reactions: old and new approaches}


\author{Vladimir Gudkov}
\email[]{gudkov@sc.edu}

\author{Young-Ho Song}
\email[]{song25@mailbox.sc.edu}
\affiliation{Department of Physics and Astronomy, University of South Carolina, Columbia, SC, 29208}

\begin{abstract}
As an attempt to find a way to resolve the discrepancy between calculated by  traditional methods  PV effects and the available set of experimental data, we analyze  the methods of calculations of symmetry violating effects in few-body nuclear systems.  The overview of methods of calculations of PV and TRIV effects in few-body neutron induced reactions is presented with the analysis of values of calculated parameters and their accuracy.
\end{abstract}

\maketitle


\section{Introduction}
\label{}

The study of parity violating (PV) and time reversal invariance violating (TRIV) effects in low energy physics is very important  for understanding  main features of the Standard model and for a search for new physics.
During the past 50 years, many calculations of
different PV and TRIV effects in nuclear physics have been done.
The methods of calculations of PV and TRIV effects are very similar,
 therefore, the reliability of prediction of TRIV effects can be justified by the accuracy of calculation of PV effects. Unfortunately,
 in the last few years, it has become clear (see, for example \cite{Zhu:2004vw,HolsteinUSC,DesplanqueUSC,RamseyMusolf:2006dz}  and references therein) that the traditional Desplanques, Donoghue, and Holstein (DDH) \cite{Desplanques1980} method for the calculation of PV effects cannot reliably describe the available experimental data.
It could be blamed on  ``wrong'' experimental data;
however, it may be that DDH approach is not adequate for the description of the set of precise experimental data because it is based on a number of models and assumptions.
Recently, a new approach based on the effective field theory (EFT) has been introduced as a model independent parameterization of PV effects (see, papers \cite{Zhu:2004vw,RamseyMusolf:2006dz} and references therein), and some calculations for two-body systems have been done \cite{Liu:2006dm}.
The power of the EFT approach could be utilized if we can analyze a large enough number of PV effects to be able to constrain all free parameters of the theory, which are usually called  low energy constants (LEC), to guarantee the adequate description of the strong interaction hadronic part of symmetry violating observables.
Then, if discrepancies between experimental data and EFT calculations will persist, it will be a clear indication that the problems are related to weak interactions in nuclei and probably to a manifestation of new physics.

Unfortunately, the number of experimentally measured (and independent in terms of unknown constants) PV effects in two body systems is not enough to practically constrain all LECs.
Therefore, one has to include into analysis few-body systems and even heavier nuclei, which are actually preferable from an experimental point of view, because usually, the measured effects in nuclei are much larger than in nucleon-nucleon scattering due to nuclear enhancement factors \cite{Sushkov:1982fa,Bunakov:1982is,Gudkov:1991qg}.
To verify the applicability of the EFT  approach for calculations of symmetry violating effects in  nuclear reactions, it is natural to start from a scattering problem in  three-body systems, and to develop a regular and self-consistent approach for calculation of symmetry violating amplitudes in a few-body systems, which later could be extended to many-body systems. From this point of view, we present an analysis of different methods of calculation of symmetry violating effects in few-body nuclear systems. Taking into account the similarity of techniques for calculation PV and TRIV effects, we will refer further only to PV effects assuming that the same could be applied for calculation of TRIV effects.

\section{Methods of Calculations  }
\label{}
Because of small value of weak coupling constants, PV effects in nuclei can be calculated in the first order of perturbation theory and  represented  as a sum of terms multiplied by corresponding  weak interaction related constants. For example, in  the DDH approach,  nuclear PV effect is represented as a linear superposition of  terms multiplied by corresponding weak
 nucleon-meson  coupling constants.
 In EFT approach, one can expect to obtain a description of nuclear PV effects as a linear superposition of terms multiplied by small numbers of LECs. To obtain PV amplitudes, we may introduce
the weak potential derived from EFT Lagrangian or directly
sum Feynman diagrams from EFT Lagrangian.
 Although the derivation of PV EFT potential would not be unique, the calculation of two body PV amplitudes can be well defined. Thus,
let us call the first approach the ''hybrid" method and the other one the ''true" EFT method.
For all these cases with potentials, the important issue is   how to calculate nuclear wave functions:  to use Schr\"{o}dinger equations or Faddeev-type few body equations. It should be noted that, due to small value of weak interactions, Distorted Wave Born Approximation (DWBA) is a standard method  for the calculation of PV amplitudes using wave-function obtained both from  Schr\"{o}dinger and Faddeev equations.
Therefore, one can classify all methods of calculations first by the method of description of two-body weak interactions: DDH or EFT methods. The DDH potential can be used to calculate amplitudes with wave functions obtained from Schr\"{o}dinger or Faddeev equations, which leads to "nuclear reaction" or "few-body equations" approaches.

\subsection{``True'' EFT approach}
\label{}
To relate  PV reaction matrices (amplitudes) for neutron nuclei interactions with two-nucleon weak reaction matrices, one can use AGS-type \cite{AGS} of Faddeev equations \cite{Faddeev:1960su} written in terms of transition operators $U_{\beta \alpha}$:
\begin{equation}\label{strong}
 U^s_{\beta \alpha}=\bar{\delta}_{\alpha \beta}G^{-1}_0 + \sum_{\gamma}\bar{\delta}_{\gamma \beta}t^s_{\gamma}G_{0}U^s_{\gamma \alpha},
\end{equation}
where upper idex $s$ indicates strong interactions, $\bar{\delta}_{\alpha \beta}=1-\delta_{\alpha \beta}$,  $t_{\gamma}$ is a two-particle transition operator in three-particle space,
 $\alpha,\beta,\dots$ are channel index,
and
\begin{equation}\label{freeres}
   G_0(z)=\frac{1}{z-H_0}
\end{equation}
is the resolvent operator for free motion.
Then, as it was shown in paper \cite{Gudkov:2010pt}, the
PV transition operator can be expressed in terms of PV two-particle  scattering operator $t^w_\gamma$ using the integral equation
\begin{equation}\label{weak}
   U^w_{\beta \alpha}=
    \sum_{\gamma}\bar{\delta}_{\gamma \beta}t^s_{\gamma}G_{0}U^w_{\gamma \alpha}+\sum_{\gamma}\bar{\delta}_{\gamma \beta}t^w_{\gamma}G_{0}U^s_{\gamma \alpha},
\end{equation}
where two-body  scattering operators  are written as a sum of PC  and PV (indicated by $w$ for weak interactions) parts:
\begin{equation}\label{texp}
   t_\gamma=t^s_\gamma +t^w_\gamma.
\end{equation}
 It should be noted that the first term of the integral equation(\ref{weak}) has exactly the same kernel as the equation (\ref{strong}) which describes the process with  strong interactions only. The second term of Eq.(\ref{weak}) does not depend on $U^w_{\beta \alpha}$ and, therefore, it is a free term in the integral equations.
 One can see that the first (integral) term includes strong two-body transition operators but the second one (free term) contains a direct contribution to $U^w_{\beta \alpha}$ from weak two-body transition operators. Therefore, Eq.(\ref{weak}) gives us a framework for a calculation of symmetry violating amplitudes using Faddeev type three-body equations in terms of two-body amplitudes, which can be obtained from EFT \cite{Zhu:2004vw,RamseyMusolf:2006dz,Liu:2006dm}.
Unfortunately, this method has not been applied to few-body systems yet.

\subsection{Faddeev equations with DDH potentials}
\label{}
The natural way to calculate PV effects in the DDH formalism  for few-body systems is to calculate  PV amplitudes using DWBA with  wave-function obtained from few-body equations (see, for example \cite{Schiavilla:2008ic,Song:2010sz}). In this case, PV effects can be represented as a linear superposition of contributions from different parts of the DDH potential weighted by corresponding weak meson-nucleon coupling constants. To see how PV effects are sensitive to  these DDH coupling constants, one can consider results of calculations of PV neutron spin rotation $\phi$ and neutron spin asymmetry $P$ in the propagation of polarized neutrons through unpolarized deuteron target. We define
 the angle of neutron spin rotation per unit length of the target sample  in terms of elastic scattering amplitudes at zero angle for opposite helicities $f_+$ and $f_-$ as
\begin{equation}
\label{eq:neutronspinrot1}
\frac{d\phi}{dz}
=-\frac{2\pi N}{p}{\rm Re}
  \left(f_{+}
  -f_{-}\right),
\end{equation}
where $N$ is a number of target nuclei per unit volume and
$p$ is a relative neutron momentum. The neutron spin asymmetry is equal to
  the relative difference of total  cross sections $P=(\sigma_+-\sigma_-)/(\sigma_++\sigma_-)$ for neutrons with opposite helicities.

 Then, one can calculate these PV effects for three different choices of DDH constants in Table \ref{tbl:LEC:DDH} which correspond the the DDH "best" values, 4-parameter fit \cite{Bowman}, and 3-parameter fit \cite{Bowman}. As a result of the calculations,  PV neutron spin rotation angles have about the same value
 $7.68$, $-6.82$ and $-8.91$
 in units of  $10^{-9}\mbox{ rad-cm}^{-1}$ for these three choices of weak coupling constants, correspondingly. The values of neutron spin asymmetry   are 
 $8.99$, $-2.09$, and $-2.36$
 in units of  $10^{-9}$. This shows that different PV parameters have, in general, different sensitivity to the DDH coupling constants. Therefore, making a large enough  number of high accuracy measurements of different PV effects, one can, in theory,  constrain values of weak coupling constants. However, it is a very difficult task because  values of PV effects in few-body nuclei usually are very small.
 \begin{table}[H] 
\caption{\label{tbl:LEC:DDH}
DDH PV coupling constants in units of $10^{-7}$.
Strong couplings are
$\frac{g^2_\pi}{4\pi}=13.9$,
$\frac{g^2_\rho}{4\pi}=0.84$,
$\frac{g^2_\omega}{4\pi}=20$,
$\kappa_\rho=3.7$, and
$\kappa_\omega=0$.
4-paramter fit and 3-parameter fit uses the same $h_\rho^1$
and $h_\omega^1$ with DDH `best'.
}
\begin{tabular}{cccc}
\hline \hline
DDH Coupling& DDH `best' & 4-parameter fit\cite{Bowman}
                         & 3-parameter fit\cite{Bowman} \\
\hline
$h^1_\pi$ &  $+4.56$        &  $-0.456$&  $-0.5$ \\
$h_\rho^0$ & $-11.4$        &  $-43.3$  & $-33$ \\
$h_\rho^2$ & $-9.5$         & $37.1$   & $41$ \\
$h_\omega^0$&  $-1.9$       & $13.7$  & $0$ \\
$h_\rho^1$ &   $-0.19$    &  $-0.19$ & $-0.19$ \\
$h_\omega^1$ & $-1.14$    &  $-1.14$ & $-1.14$ \\
\hline \hline
\end{tabular}
\end{table}

\subsection{``Hybrid'' EFT approach}
\label{}
The "hybrid" approach uses the same calculation scheme,  as it was described in the previous section, but instead of using DDH potentials, it uses the potentials derived from a particular choice of EFT Lagrangian. However, as it was shown in \cite{Schiavilla:2008ic},   the potentials derived from pionless EFT lagrangian \cite{Zhu:2004vw} contain contributions from the same operators  as the DDH potentials \cite{Desplanques:1979hn}. This is also correct for potentials derived from pionful EFT Lagrangian \cite{Zhu:2004vw}. Therefore, all these potentials (DDH, EFT pionless and EFT pionful) can be expanded in terms of a set of $O^{(n)}_{ij}$ operators   as
\begin{equation}
v_{ij}^\alpha=\sum_{n} c_n^\alpha O^{(n)}_{ij},\quad
\mbox{$\alpha=$ DDH or pionless EFT or pionful EFT}
\end{equation}
where
 operators $O^{(n)}_{ij}$ are defined as  products
 of isospin, spin,  and
vector operators ${\bm X}^{(n)}_{ij,\pm}$
\bea
{\bm X}^{(n)}_{ij,+} & \equiv&[\vp_{ij},f_n(r_{ij})]_{+},\no
{\bm X}^{(n)}_{ij,-}&\equiv&i [\vp_{ij},f_n(r_{ij})]_{-},
\eea
with $\vp_{ij}=(\vp_i-\vp_j)/{2}$, and  $c_n^\alpha$ being expansion parameters.
The parameters of this expansion for DDH, EFT pionless, and EFT pionful potentials are presented in   Table \ref{tbl:pvpotential} (see paper \cite{Song:2010sz}  for more details and discussions).
\begin{table}[H] 
\caption{\label{tbl:pvpotential}
 Parameters and operators of parity violating potentials.
 $\pi NN$ coupling $g_{\pi NN}$ can be represented by $g_A$ by using
Goldberger-Treiman relation, $g_\pi=g_A m_N/ F_\pi$ with $F_\pi=92.4$ MeV.
${\cal T}_{ij}\equiv (3\tau_i^z\tau_j^z-\tau_i\cdot\tau_j)$.
}
\begin{tabular}{cccccccc}
\hline \hline
$n$ & $c_n^{DDH}$ & $f_n^{DDH}(r)$ & $c_n^{\not{\pi}}$ & $f_n^{\not{\pi}}(r)$ & $c_n^{\pi}$ & $f_n^{\pi}(r)$ & $O^{(n)}_{ij}$ \\
\hline
$1$ & $+\frac{g_\pi }{2\sqrt{2} m_N}h_\pi^1$ & $f_\pi(r)$ &
      $\frac{2\mu^2}{\Lambda^3_\chi}   C^{\not{\pi}}_6$ &
      $f^{\not{\pi}}_\mu(r)$ &
      $+\frac{g_\pi }{2\sqrt{2} m_N}h_\pi^1$ & $f_\pi(r)$ &
      $(\tau_i \times \tau_j )^z (\vs_i + \vs_j ) \cdot {\bm X}^{(1)}_{ij,-}$
\\
$2 $ & $ -\frac{g_\rho}{m_N}h_\rho^0 $ & $ f_\rho(r) $ & $
      0 $ & $ 0 $ & $
      0 $ & $ 0 $ & $
      (\tau_i\cdot\tau_j)(\vs_i-\vs_j)\cdot{\bm X}^{(2)}_{ij,+}$
\\
$3 $ & $ -\frac{g_\rho(1+\kappa_\rho)}{m_N} h_\rho^0 $ & $f_\rho(r) $ & $
      0 $ & $ 0$ & $
       0 $ & $ 0 $ & $
      (\tau_i\cdot\tau_j)(\vs_i\times\vs_j)\cdot{\bm X}^{(3)}_{ij,-}$
\\
$4 $ & $ -\frac{g_\rho}{2 m_N} h_\rho^1 $ & $ f_\rho(r) $ & $
      \frac{\mu^2}{\Lambda^3_\chi}(C^{\not{\pi}}_2+C^{\not{\pi}}_4)
      $ & $ f_\mu^{\not{\pi}}(r) $ & $
      \frac{\Lambda^2}{\Lambda^3_\chi}(C^{\pi}_2+C^{{\pi}}_4) $ & $
      f_\Lambda(r) $ & $
      (\tau_i+\tau_j)^z(\vs_i-\vs_j)\cdot{\bm X}^{(4)}_{ij,+}$
\\
$5  $ & $  -\frac{g_\rho(1+\kappa_\rho)}{2 m_N}h_\rho^1 $ & $ f_\rho(r)
     $ & $ 0 $ & $ 0 $ & $
     \frac{2\sqrt{2}\pi g_A^3\Lambda^2}{\Lambda_\chi^3}h^1_\pi $ & $
      L_\Lambda(r) $ & $
      (\tau_i+\tau_j)^z(\vs_i\times\vs_j)\cdot{\bm X}^{(5)}_{ij,-}$
\\
$6 $ & $ -\frac{g_\rho}{2\sqrt{6} m_N}h_\rho^2 $ & $ f_\rho(r) $ & $
  -\frac{2\mu^2}{\Lambda^3_\chi}C_5^{\not{\pi}} $
  & $   f^{\not{\pi}}_\mu(r)$ & $
  -\frac{2\Lambda^2}{\Lambda^3_\chi}C_5^{{\pi}} $
  & $ f_\Lambda(r)$ & $
   {\cal T}_{ij}
   (\vs_i-\vs_j)\cdot{\bm X}^{(6)}_{ij,+}$
\\
$7 $ & $ -\frac{g_\rho(1+\kappa_\rho)}{2\sqrt{6} m_N}h_\rho^2 $ & $ f_\rho(r) $ & $
   0 $ & $ 0 $ & $
    0  $ & $ 0 $ & $
   {\cal T}_{ij}(\vs_i\times\vs_j)\cdot{\bm X}^{(7)}_{ij,-}$
\\
$8 $ & $ -\frac{g_\omega}{m_N}h_\omega^0 $ & $ f_\omega(r) $ & $
   \frac{2\mu^2}{\Lambda^3_\chi} C_1^{\not{\pi}} $ & $ f_\mu^{\not{\pi}}(r) $ & $
   \frac{2\Lambda^2}{\Lambda^3_\chi} C_1^{{\pi}} $ & $ f_\Lambda(r) $ & $
      (\vs_i-\vs_j)\cdot{\bm X}^{(8)}_{ij,+}$
\\
$9  $ & $  -\frac{g_\omega(1+\kappa_\omega)}{m_N} h_\omega^0 $ & $ f_\omega(r) $ & $
   \frac{2\mu^2}{\Lambda_\chi^3}\tilde{C}^{\not{\pi}}_1 $ & $ f_\mu^{\not{\pi}}(r) $ & $
   \frac{2\Lambda^2}{\Lambda_\chi^3}\tilde{C}^{{\pi}}_1 $ & $ f_\Lambda(r) $ & $
   (\vs_i\times\vs_j)\cdot{\bm X}^{(9)}_{ij,-}$
\\
$10 $ & $ -\frac{g_\omega}{2 m_N} h_\omega^1 $ & $ f_\omega(r) $ & $
     0 $ & $ 0 $ & $
      0  $ & $ 0  $ & $
     (\tau_i+\tau_j)^z(\vs_i-\vs_j)\cdot{\bm X}^{(10)}_{ij,+}$
\\
$11 $ & $ -\frac{g_\omega(1+\kappa_\omega)}{2m_N} h^1_\omega $ & $ f_\omega(r) $ & $
    0 $ & $  0 $ & $
     0  $ & $  0  $ & $
    (\tau_i+\tau_j)^z(\vs_i\times\vs_j)\cdot{\bm X}^{(11)}_{ij,-}$
\\
$12 $ & $ -\frac{g_\omega h_\omega^1-g_\rho h_\rho^1}{2m_N} $ & $ f_\rho(r) $ & $
   0 $ & $ 0 $ & $
    0 $ & $ 0   $ & $
   (\tau_i-\tau_j)^z(\vs_i+\vs_j)\cdot{\bm X}^{(12)}_{ij,+}$
\\
$13 $ & $ -\frac{g_\rho}{2m_N} h^{'1}_\rho $ & $ f_\rho(r) $ & $
      0 $ & $ 0 $ & $
       -\frac{\sqrt{2}\pi g_A\Lambda^2}{\Lambda_\chi^3} h_\pi^1 $ & $ L_\Lambda(r) $ & $
     (\tau_i\times\tau_j)^z(\vs_i+\vs_j)\cdot{\bm X}^{(13)}_{ij,-}$
\\
$14$ & 0  & 0 &  0 & 0 &
 $\frac{2\Lambda^2}{\Lambda^3_\chi}   C^{{\pi}}_6$ &
 $f_\Lambda(r)$ &
 $(\tau_i\times\tau_j)^z(\vs_i+\vs_j)\cdot{\bm X}^{(14)}_{ij,-}$
\\
$15$   &  0  &  0   & 0 &  0   &
   $\frac{\sqrt{2}\pi g^3_A\Lambda^2}{ \Lambda_\chi^3} h_\pi^1 $ &
   $\tilde{L}_\Lambda(r)$ &
   $(\tau_i\times\tau_j)^z(\vs_i+\vs_j)\cdot{\bm X}^{(15)}_{ij,-}$ \\
\hline \hline
\end{tabular}
\end{table}
One can see that all these potentials (DDH and EFT ones) have phenomenological radial dependencies ($f_{x}(r)$, $\tilde{L}_\Lambda(r)$, $L_\Lambda(r)$, and $H_\Lambda(r) $ ) which are related to masses of exchange mesons for the DDH case and to cutoff parameters in the case of EFT potentials (for detailed notations and discussions of the properties of these radial functions, see \cite{Song:2010sz}). Therefore, the results of calculations of PV effects in the "hybrid" approach can be represented as a linear combination of contributions from different operators weighted by (unknown) LECs and folded (integrated out) parameters which are  depended on these radial functions. This is very similar (see Table \ref{tbl:pvpotential}) to the   representation of PV effects of the DDH-type of calculation. The only practical difference of the "hybrid" approach, as compare to the DDH one,  is related to the different numbers of operators corresponding to different models of EFT. For example,  for DDH potentials, it could be up to 13 operators, while for considered EFT potentials,  it could be 5 or 9 operators for pionless and pionful EFTs, correspondingly.

 Despite the similarity of structure between DDH and ''hybrid" EFT potentials, DDH potential has additional assumptions and constraints on the forms and relations between operators.  Because the systematic fitting between ''hybrid" EFT calculation and experimental data has not been done yet, we cannot make any prediction using ''hybrid" EFT. Also the adoption of ''hybrid" EFT may bring some undesirable artificial effects. Thus, ''true" EFT calculation may be an ideal approach to the few-body PV observables.

\subsection{Nuclear reaction approach}
\label{}
 Since PV effects in a few-body systems are usually too small to expect many experimental results and precise calculations of these effects are rather difficult, it is desirable to have  a method for a reliable estimate   of possible observable parameters.  It is possible to calculate some PV effects  in the first order of perturbation theory with wave functions obtained from  a solution of Schr\"{o}dinger equation for reliable nuclear models (see, for example \cite{Avishai:1982qf,Dmitriev:1983mg}). We consider here a nuclear reaction approach which gives  PV amplitudes in DWBA with wave functions obtained from a reliable nuclear models with  parameters fixed from available experimental data.
 This  gives the opportunity to explore many nuclei and observables  and to find nuclei with reasonably large PV effects for further  experiments and detailed calculations.
To illustrate this approach, let us estimate PV effects in $n + ^3He \rightarrow ^3H + p$ reaction with polarized neutrons. The $^3He$ and $^4He$ systems were  subjects of intensive investigations  for a long time, and as a result,  many parameters related to reactions with neutrons and protons, as well as to excitation energy levels of these nuclei, have been measured and evaluated by a number of different groups. This rather comprehensive data  provides the opportunity to estimate  values of possible PV effects and their dependence on neutron energy  in $n + ^3He \rightarrow ^3H + p$ reaction using microscopic nuclear reaction theory approach (see \cite{Gudkov:He2010}).

Recently, it has been proposed
to measure PV asymmetry of protons in $n + ^3He \rightarrow ^3H + p$ reaction with polarized neutrons at the Spallation Neutron Source at the Oak Ridge National Laboratory. For typical neutron energy $E_n \sim 0.01\,eV$, which corresponds to a wave vector $k_n \sim 2.19\cdot 10^{-5}\, fm^{-1}$, the energy of outgoing protons and proton wave vector are $E_p = 0.764 \, MeV$ and $k_p = 0.19\, fm^{-1}$, correspondingly. Taking a characteristic $^3He$ radius as $R = 1.97\, fm$, one obtains  $(k_nR)\sim 4\cdot 10^{-4}$ and $(k_pR)\sim 0.4$. Therefore, for the initial channel,  contributions from $p$-wave neutrons to a reaction matrix (amplitude) are highly suppressed, whereas for the final channel, the amplitude with orbital momenta  of protons $l=0$ and $l=1$ has  the same order of magnitude. The contribution from $d$-wave protons is suppressed by a factor $\sim 0.025$; therefore, one can ignore $d$-waves  within the accuracy of our estimates.

 PV asymmetry  $\alpha _{PV}$ of outgoing protons, in  directions along to neutron polarization and  opposite to it, is proportional to  PV correlation $(\vec{\sigma}\cdot \vec{k}_p)$.
Using standard techniques (see, for example \cite{BG_NP82}), one can represent these asymmetries in terms of matrix $\hat{R}$ which is related to reaction matrix $\hat{T}$ and to $S$-matrix as
\begin{equation}\label{RTS}
   \hat{R}=2\pi i\hat{T}=\hat{1}-\hat{S}.
\end{equation}
Then, for our case,
\begin{eqnarray}\label{pv}\nonumber
 {\alpha _{PV}} &=& \frac{2}{{r}}{\mathop{\rm Re}\nolimits} [ - 3\sqrt 2  < 01|{R^1}|10 >  \cdot  < 00|{R^0}|00{ > ^*}   \\
 &+&(\sqrt 6  < 11|{R^0}|00 >  + 6 < 11|{R^1}|10 > ) < 10|{R^1}|10{ > ^*}],
 \end{eqnarray}
where
 \begin{equation}\label{dn}
    {r} = ( |< 00|{R^0}|00 >|^2   + 3 |< 10|{R^1}|10 >|^2  ).
 \end{equation}
We use spin-channel representation, where for the matrix element $< s^{\prime} l^{\prime}|{R^J}|sl >$, $l$ and $l^{\prime}$ are orbital momenta of initial and final channels with corresponding spin-channels $s$ and $s^{\prime}$ and $J$ is the total spin of the system.
Calculations of  matrix elements $< s^{\prime} l^{\prime}|{R^J}|sl >$ for PV effects in nuclear reactions have been done \cite{BG_NP82} using distorted wave Born approximation  in microscopic theory of nuclear reactions \cite{MW}.  They lead to
the PV amplitudes induced by parity violating potential $W$,
\begin{equation}
R^{fi}_{PV} = 2\pi i<{\Psi^-_f}|W|{\Psi^+_i}>,
\end{equation}
 where $\Psi^{\pm}_{i,f}$ are the eigenfunctions of the nuclear $P$-invariant
 Hamiltonian with  the  appropriate boundary conditions \cite{MW}:
\begin{equation}
 \Psi^{\pm}_{i,f}=\sum_k a^\pm_{k(i,f)}(E)\; \phi_k + \sum_m\int
b^{\pm}_{m(i,f)}(E,E')\; \chi^{\pm}_m(E')\; dE'.
 \label{eq:wf}
\end{equation}
Here, $\phi_k$ is the wave function of  the $k^{th}$
  resonance and $\chi^{\pm}_m(E)$ is the potential
 scattering wave function in the channel $m$.
The coefficient
\begin{equation}
 a^\pm_{k(i,f)}(E)={\exp{(\pm i\delta_{i,f})}\over {(2\pi)^{1\over 2}}}{{(\Gamma^{i,f}_k)^{1\over 2}}\over {E-E_k\pm{i\over
   2}\Gamma_k}}
\end{equation}
describes  nuclear resonances contributions and the
coefficient $b^{\pm}_{m(i,f)}(E,E')$ describes potential
scattering and interactions between the continuous spectrum and
 resonances. Here, $E_k$, $\Gamma_k$, and $\Gamma^i_k$ are
the energy, the total width, and the partial width in the channel
$i$ of the $k$-th  resonance, $E$ is the neutron
energy, and $\delta_i$ is the potential scattering phase in the
channel $i$; $(\Gamma^i_k)^{1\over 2} = (2\pi )^{1\over 2}
 <{\chi_i(E)}|V|{\phi_k}>$,
 where $V$ is a residual interaction operator.
 As  was shown in \cite{BG_NP82} for nuclei with rather large atomic numbers, the resonance contribution  is  dominant. Then, for the simplest case with only two resonances with opposite parities,  the expressions for matrix element  $\hat{R}$ for neutron-proton reaction  with parity violation is:
 \begin{equation}
< s^{\prime} l^{\prime}|{R^J}|sl > = -
{{iw(\Gamma^n_l(s)\Gamma^p_{l^{\prime}}(s^{\prime}))^{1\over
2}}\over{(E-E_l+i\Gamma_l/2)(E-E_{l^{\prime}}+i\Gamma_{l^{\prime}}/2)}}{\it
e}^{i(\delta^n_l + \delta^p_{l^{\prime}})}, \label{eq:pv}
\end{equation}
and with conservation of parity (one resonance contribution) is:
\begin{equation}
< s^{\prime} l^{\prime}|{R^J}|sl > =
{{i(\Gamma^n_l(s)\Gamma^p_{l^{\prime}}(s^{\prime}))^{1\over
2}}\over{(E-E_l+i\Gamma_l/2)}}{\it
e}^{i(\delta^n_l + \delta^p_{l^{\prime}})}, \label{eq:pc}
\end{equation}
where $w=-\int \phi_l W \phi_{l^{\prime}}d\tau$ is PV nuclear matrix element mixing parities of two resonances.

This technique has been proven to work very well for the calculation of nuclear PV effects for intermediate and heavy nuclei. Assuming the dominant resonance contribution to PV effects for the $n + ^3He \rightarrow ^3H + p$ reaction, we use  this approach  to estimate  characteristic values of PV  effects  using parametrization of PV effects in terms of known  resonance structure of the system. Fortunately,  the detailed structure of resonances ($^4He$ levels) \cite{Tilley:1992} and low energy neutron scattering parameters \cite{Mughabghab} are well known for this reaction from numerous experiments.

To estimate PV  asymmetry in  $n + ^3He \rightarrow ^3H + p$ reactions using the described formalism, we  take into account all known resonances \cite{Tilley:1992,Mughabghab} which result in multi-resonance representation for $\hat{R}$ matrix elements. From the selection rules for angular momenta (see Eq.(\ref{pv}) and general expressions in \cite{BG_NP82}), one can see that  for low energy neutrons only resonances with the total spin of $J=0,1$  contribute to PV asymmetries of the interest.  Thus, we consider contributions from nine low energy resonances \cite{Tilley:1992,Mughabghab} (see Table (\ref{tab_res1}) ): one resonance with total angular momentum and parity $J^{\pi}=0^+$, three with $J^{\pi}=0^-$, four with $J^{\pi}=1^-$, and one with $J^{\pi}=1^+$. For further calculations, we assume that all weak matrix elements, which mix resonances with opposite parities,   have the same values and are described by a phenomenological formula \cite{BG_NP82} $w=2\cdot10^{-4}eV\sqrt{\bar{D}(eV)}$ (where  $\bar{D}$ is an average energy level spacing). This formula is in good agreement with other statistical nuclear model estimates \cite{Kadmensky:1983,Bunakov:1989,Johnson:1991} of nuclear weak matrix elements for medium and heavy nuclei. The extrapolation of this formula to the region of one-particle nuclear excitation  leads to the correct  value for weak nucleon-nucleon interaction. Therefore, one can use this approximation for  rough estimates  of average values of weak matrix elements in few-body systems. This leads to the value of weak matrix element $w=0.5\;eV$ (with $\bar{D}\simeq 6\; MeV$),  which is rather close to the typical value of one particle weak matrix element. One can see from Eqs.(\ref{eq:pv}) and (\ref{eq:pc}) that the expressions for PV and PC $\hat{R}$ matrices depend not on  neutron and proton partial widths   but on their  amplitudes,  the values of which depend on particular spin-channels. Since we know only partial widths, we have to make  assumptions about values of amplitudes of partial widths for a specific spin-channel and about their signs (phases).   This leads to another uncertainty in our estimation in addition to the previously  given  assumption about weak matrix elements.   To treat the spin-channel dependence of partial width amplitudes, we assume that  partial widths for each spin-channel are equal to each other.  This gives us an average factor of uncertainly of about $2$. The signs of  width amplitudes, as well as the signs of weak matrix elements $w$, are left undetermined (random). This also can lead to a factor of uncertainly of $2$ or $3$. Therefore, one can see that the uncertainly of our multi-resonance calculations is about of one order of magnitude.

Taking into account these considerations  and using resonance parameters \cite{Tilley:1992,Mughabghab} of the Table (\ref{tab_res1}), one can estimate the PV asymmetry for thermal neutrons   as
\begin{equation}\label{pv1}
    {\alpha _{PV}}=-(1-4)\cdot 10^{-7}.
\end{equation}
The set of resonance parameters of the Table (\ref{tab_res2}) results in a slightly lager   PV asymmetry
\begin{equation}\label{pv2}
    {\alpha _{PV}}=-(4-8)\cdot 10^{-7}.
\end{equation}
The difference between these two sets  is related to the discrepancy between \cite{Mughabghab} and \cite{Tilley:1992} for resonance parameters for the first positive resonance ($E_n=0.430\; MeV$).

To show  contributions of each resonance to PV asymmetry $\alpha _{PV}$, we normalized contributions from each resonance in terms of relative intensity  to the  strongest one, which is taken as $100\%$ (see last two columns in  Tables (\ref{tab_res1}) and (\ref{tab_res2})).  Some resonances contribute through two different spin-channels, $s=0$ and $s=1$. In those cases, the contributions from two spin-channels can be either with the same sign or with the opposite sign, depending on unknown phases of amplitudes of partial widths and weak matrix elements (see, for example, resonance at $3.062\; MeV$ in Table (\ref{tab_res1})). As can be seen from these tables,   different resonances contribute essentially differently to the value of PV violating effects.
Moreover,  different sets of resonance parameters can change  weights of the resonances for a particular asymmetry. For example, the lowest $0^-$-resonance contribution to  the asymmetry ${\alpha _{PV}}$ appears to be 3\% using the parameters of Table (\ref{tab_res1}), while it would be the dominant one using the parameters of Table (\ref{tab_res2}). This is related to the fact that for the set of Table (\ref{tab_res1}), the contribution of the $0^-$-resonance to the ${\alpha _{PV}}$ is suppressed by a factor of about 40 due to destructive interference between parity conserving and parity violating amplitudes.
Therefore, the readability of this method can be essentially improved by increasing the accuracy in measurements of parameters of the most ``important'' resonances.

It should be noted that the estimated value of the PV asymmetry $\alpha _{PV}$ at thermal energy (see Eqs. (\ref{pv1}) and (\ref{pv2})) is surprisingly in very good agreement with exact calculations for zero energy neutrons \cite{Viviani:2010qt}. This could be considered as an additional argument for reliability of the suggested resonance approach. Also, matching the estimated value of the observable parameter with exact calculations at low energy gives us the opportunity to predict PV effects in a wide range of neutron energies.

\begin{table}[H]
 \caption{Resonance parameters (Set 1). Here, $E_r$ is a resonance energy; $T$ and $J^{\pi}$  are resonance isospin and the total resonance spin with parity; $\Gamma$ and $\Gamma_p$ are total and proton widths; $\Gamma_n$, $\Gamma^0_n$ and  $l$ are neutron width, reduced width, and angular momentum, correspondingly; and $\alpha _{PV}$ (\% ) is normalized contribution of the resonance to  $\alpha _{PV}$. \label{tab_res1}}
\begin{tabular}{| c | c | c | c | c | c | c | c | c | }
 \hline
  $E_r (MeV)$&$\;J^{\pi}\;$&$\;\; l\;\; $&$\;\;T\;\;$& $\Gamma_n(MeV)$ & $\Gamma^0_n(eV)$ &  $\Gamma_p(MeV)$ &  $\Gamma(MeV)$ & $\alpha _{PV}$ (\% ) \\
 \hline
-0.211 & 0+ & 0 & 0 &   & 954.4 & 1.153 & 1.153 &  \\
0.430 & 0- & 1 & 0 & 0.48  &   & 0.05 & 0.53 & 3.1 \\
3.062 & 1- & 1 & 1 & 2.76  &   & 3.44 & 6.20 & 100$\pm$26 \\
3.672 & 1- & 1 & 0 & 2.87  &   & 3.08 & 6.10 & 75$\pm$24 \\
4.702 & 0- & 1 & 1 & 3.85  &   & 4.12 & 7.97 & 20 \\
5.372 & 1- & 1 & 1 & 6.14  &   & 6.52 & 12.66 & 79$\pm$18 \\
7.732 & 1+ & 0 & 0 & 4.66  &   & 4.725 & 9.89 &  \\
7.792 & 1- & 1 & 0 & 0.08  &   & 0.07 & 3.92 & 2$\pm$1  \\
8.062 & 0- & 1 & 0 & 0.01  &   & 0.01 & 4.89 & 14 \\
  \hline
\end{tabular}
 \end{table}

\begin{table}[H]
 \caption{Resonance parameters (Set 2). Here, $E_r$ is a resonance energy; $T$ and $J^{\pi}$  are resonance isospin and the total resonance spin with parity; $\Gamma$ and $\Gamma_p$ are total and proton widths; $\Gamma_n$, $\Gamma^0_n$ and  $l$ are neutron width, reduced width, and angular momentum, correspondingly;  $\alpha _{PV}$ (\% ) is normalized contribution of the resonance to  $\alpha _{PV}$.\label{tab_res2}}
\begin{tabular}{| c | c | c | c | c | c | c | c | c | }
 \hline
  $E_r (MeV)$&$\;J^{\pi}\;$&$\;\; l\;\; $&$\;\;T\;\;$& $\Gamma_n(MeV)$ & $\Gamma^0_n(eV)$ &  $\Gamma_p(MeV)$ &  $\Gamma(MeV)$ & $\alpha _{PV}$ (\% ) \\
 \hline
-0.211 & 0+ & 0 & 0 &   & 954.4 & 1.153 & 1.153 &    \\
0.430 & 0- & 1 & 0 & 0.20  &   & 0.640 & 0.84 & 100 \\
3.062 & 1- & 1 & 1 & 2.76  &   & 3.44 & 6.20 & 82$\pm$27 \\
3.672 & 1- & 1 & 0 & 2.87  &   & 3.08 & 6.10 & 62$\pm$20 \\
4.702 & 0- & 1 & 1 & 3.85  &   & 4.12 & 7.97 & 16 \\
5.372 & 1- & 1 & 1 & 6.14  &   & 6.52 & 12.66 & 65$\pm$15 \\
7.732 & 1+ & 0 & 0 & 4.66  &   & 4.725 & 9.89 &   \\
7.792 & 1- & 1 & 0 & 0.08  &   & 0.07 & 3.92 & 2$\pm$1 \\
8.062 & 0- & 1 & 0 & 0.01  &   & 0.01 & 4.89 & 1 \\
  \hline
\end{tabular}
 \end{table}

  \begin{figure}
\includegraphics{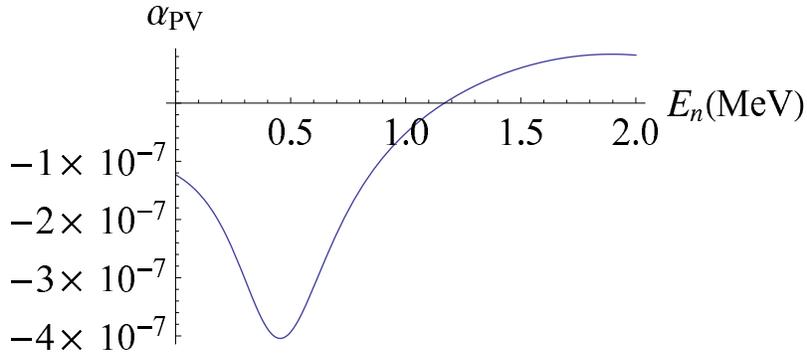}
\caption{(Color online) Resonance enhancement of of the $\alpha _{PV}$ asymmetry (for the first set of parameters). \label{a1plot}}
 \end{figure}

\section{Conclusions}
\label{}
 Considering a variety of methods of the calculation of PV effects in low energy
few-body nuclear reactions, the EFT approach can be a solution for the discrepancy between DDH description of PV effects and experiments. Furthermore, the ''true'' EFT approach, which involves a solution of AGS-type of few body equations, can be the most promising method without introducing any additional model dependencies or parameters.

At the same time, the use of all other approaches is still useful in some cases. In particular, the simple nuclear reaction approach could be very useful for a preliminary study of new PV effects in few-body system, since it does not require heavy  calculations and provides estimates for both PV and parity conserving effects for rather wide energy regions.
\\

{\bf Acknowledgments}
This work was supported by the DOE grants no. DE-FG02-09ER41621.
\\





\bibliography{ParityViolation}







\end{document}